\documentclass[conference,a4paper]{IEEEtran}
\usepackage{epsfig,  subfigure}
\usepackage[latin1]{inputenc}
\usepackage{subeqnarray}
\usepackage{amsfonts}
\usepackage{amsmath}
\usepackage{amssymb}
\usepackage{graphicx}
\usepackage{color}
\title{Multimodal Approach for Video Surveillance Indexing and Retrieval}

\author{Ali WALI and Adel M. ALIMI \\
\emph{REGIM: REsearch Groups on Intelligent Machines, University of Sfax,}\\
\emph{National School of Engineers (ENIS), BP 1173, Sfax, 3038, Tunisia}\\
\emph{ali.wali@ieee.org, adel.alimi@ieee.org}\\}
\begin{document}
\maketitle
\begin{abstract}
In this paper, we present an overview of a new approach to indexing and searching the video sequence by the content that has been developed within the REGIMVid project.
The platform termed MAVSIR provides High-level feature extraction from audio-visual content and concept/event-based video retrieval.  We describe the architecture of the system as well as provide an overview of the descriptors supported to date. We then demonstrate the usefulness of the toolbox in the context of feature extraction, concepts/events learning and retrieval in large collections of video surveillance dataset.
\end{abstract}
\section{Introduction}
\label{Sec:int}
Image and video indexing and retrieval continue to be an
extremely active area within the broader multimedia research
community \cite{trec06,trec07}. Interest is motivated by the very real 
requirement for efficient techniques for indexing large archives of
audiovisual content in ways that facilitate subsequent usercentric
accessing. Such a requirement
is a by-product of the decreasing cost of storage and the now
ubiquitous nature of capture devices. The result of which is
that content repositories, either in the commercial domain
(e.g. broadcasters or content providers repositories) or the
personal archives are growing in number and size at virtually exponential rates.
It is generally acknowledged that providing truly efficient usercentric access to large content archives requires
indexing of the content in terms of the real world semantics
of what it represents.

 Furthermore, it is acknowledged
that real progress in addressing this challenging task requires
key advances in many complementary research areas such as;
scalable coding of both audiovisual content
and its metadata, database technology and user interface design. 
The REGIMVid project integrates many of these issues.In figure 1 we present our REGIMVid subsystem. A
key effort within the project is to link audio-visual analysis
with concept reasoning in order to extract semantic
information. In this context, high-level pre-processing
is necessary in order to extract descriptors that can be subsequently
linked to the concept and used in the reasoning
process.
\begin{figure}[htb]
  \centerline{\epsfig{figure=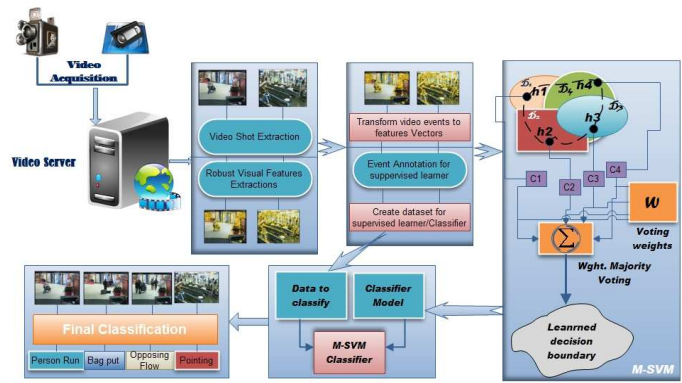,width=9cm}}
  \caption{Our REGIMVid subsystem Architecture}
	\label{fig:Drawing1}
  \end{figure}
 In addition to concept-based reasoning,
the project has other research activities that require high-level
feature extraction (e.g. semantic summary of metadata
\cite{M.Ellouze06}, Text-based video retrieval \cite{Karray.H.07, Wali.A07}, event detection \cite{bib3} and Semantic Access to Multimedia Data \cite{bib10}) it was
decided to develop a common platform for descriptor extraction
that could be used throughout the project.
In this paper, we describe our subsystem for video surveillance indexing and retrieval. The remainder of the paper is organised
as follows: a general overview of the toolbox is provided
in Section 2, include a description of the architecture. In section 3 we present our approach to detect and extract of moving objects from video surveillance dataset. It includes a presentation of different concepts taken care by our system.We present the combining single SVM classifier for learning video event/concept in section 4. The descriptors of the visual feature extraction will be presented in section 5. Finally, we present our experimental results for both event and concept detection future plans for both the extension of the toolbox and its use in different scenarios.

\section{MAVSIR TOOLBOX for video surveillance indexing Overview}
In this section, we present an overview of the structure of
the toolbox. The
MAVSIR Toolbox currently supports extraction of 10 low-level (see section 5)
visual descriptors. The design is based on the architecture
of the MPEG-7 eXperimentation Model (XM),
the official reference software of the ISO/IEC MPEG-7 standard. 

The main objective of our system is to provide automatic content analysis using concept/event-based and low-level features. The system (figure 2) first detect and segment the moving object from video surveillance dataset. In the second step, it extracts three class of features from the each frame, from a static background and the segmented objects(the first class from $\Omega_{in}$ , the second from $\Omega_{out}$ and the last class is from each key-frame in RGB color space,see subsection 3.2), and labels them based on corresponding features. For example, if three features are used (color, texture and shape), each frame has at least three labels from $\Omega_{out}$, three labels from $\Omega_{in}$ and three labels from key-frame.

This reduces the video as a sequence of labels containing the common features between consecutive frames. The sequence of labels aim to preserve the semantic content, while reducing the video into a simple form. It is apparent that the amount of data needed to encode the labels is an order of magnitude lower than the amount needed to encode the video itself. This simple form allows the machine learning techniques such as Support Vector Machines to extract high-level features. 
\begin{figure}[htb]
  \centerline{\epsfig{figure=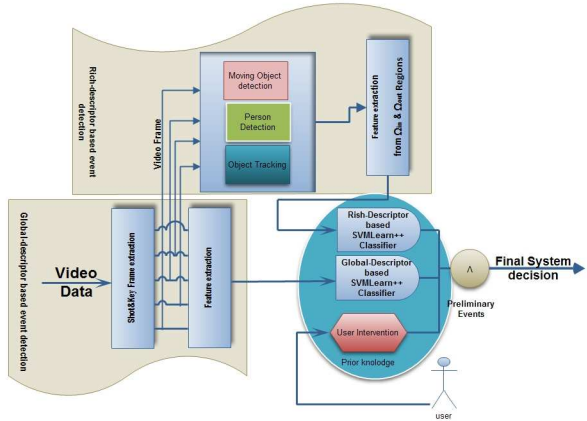,width=10cm}}
  \caption{Overview of our system for video input}
	\label{fig:Drawing1}
  \end{figure}

Our method offer a way to combine low-level features witch enhances the system performance. The high-level features extraction system according to our toolkit provides an open framework that allows easy integration of new features. In addition, the Toolbox can be integrated with traditional methods of video analysis. Our system offers many functionalities at different granularity that can be applied to applications with different requirements. The Toolbox also provides a flexible system for navigation and display using the low-level features or their combinations. Finally, the feature extraction according to the Toolbox can be performed in the compressed domain and preferably real-time system performance such as the videosurveillance systems. 
\section{Moving object detection and extraction}
To detect and extract a moving object from a video dataset we use a region-based active contours model where the designed objective function is composed of a region-based term and optimize the curve position with respect to motion and intensity properties. The main novelty of our approach is that we deal with the motion estimation by optical flow computation and the tracking problem simultaneously. Besides, the active contours model is implemented using a level set, inspired from Chan and Vese approach \cite{Chan1999}, where topological changes are naturally handled.
\subsection{Motion estimation by optical flow} 
In our system, we use gradient-based optical flow algorithm proposed by Horn and Schunck \cite{horn81}. similar to T. Macan and S. Loncaric \cite{ST2001},we have integrated the algorithm in multi-grid technique where the image is decomposed into Gaussian pyramid-set of the reduced images. The calculation starts at a coarser scale of the image decomposition, and the results are propagated to finer scales.  \\
Let us suppose that the intensity of the image at a time t and position (x, y) is given by I (x, y, t). The assumption on brightness constancy is made that the total derivative of brightness function is zero which results the following equation:
\begin{equation}
\frac{\partial I}{\partial x}\frac{dx}{dt}+\frac{\partial I}{\partial y}\frac{dy}{dt}+\frac{\partial I}{\partial t}=0\;or\;I_{x,i}u_{i}+I_{y,i}v_{i}+I_{t,i}=0
\end{equation}
This equation is named 'Brightness Change Constraint Equation'. Where u and v are components of optical flow in horizontal and vertical directions, respectively, and $I_{x}$, $I_{y}$ and $I_{t}$ are partial derivatives of I with respect to x, y and t respectively. Horn and Schunck added additional smoothness constraint because the equation (1) is insufficient to compute both components of optical flow. They minimized weighted sum of smoothness term and brightness constraint term:
\begin{equation}
\int_{\Omega}(I_{x}u+I_{y}v+It)^{2}+\lambda(\left\|\nabla{u}\right\|^{2}+\left\|\nabla{v}\right\|^{2})dx 
\end{equation}
Minimization and discretization of equation (2) results in two equations for each image point where vector values $u_i$ and $v_i$ are optical flow variables to be determined. To solve this system of differential equations, we use the iterative Gauss-Seidel relaxation method (for more detail see $http://benallal.free.fr/an/Optim6/Optim6.htm$).
\subsection{Our moving object segmentation model}
In our case, taking into consideration the motion information obtained by calculating the optical flow, we propose the following descriptors for the segmentation of mobile objects in a video surveillance dataset: \\
\begin{equation}
\begin{array}{l}
k_{in}(x,\Omega_{in})=\lambda\left|SV_{g}(x)-c_{1}(\Omega_{in})\right|^{2}\\
k_{out}(x,\Omega_{out})=\lambda\left|SV_{g}(x)-c_{2}(\Omega_{out})\right|^{2}\\
k_{b}(x)=\mu
\end{array}
\end{equation}
With $c_{1}$ is the average of the region $\Omega_{in}$, $c_2$ is the average of the region $\Omega_{out}$,$\mu$ and $\lambda$ constants positive. SVg(x) is the image obtained after a threshold of the optical flow velocity and applicate of a gaussian filter.
The values of $c_{1}$ and $c_{2}$ are re-estimated during the spread of the curve. The method of levels sets is used directly representing the curve $\Gamma(x)$ as the curve of zero to a continuous function U(x). Regions and contour are expressed as follows:\\
\begin{equation}
\begin{array}{l}
\Gamma=\partial\Omega_{in}=\left\{x\in\Omega_{I}/U(x)=0\right\}\\
\Omega_{in}=\left\{x\in\Omega_{I}/U(x)<0\right\}\\
\Omega_{out}=\left\{x\in\Omega_{I}/U(x)>0\right\}\\
\end{array}
\end{equation}
The unknown sought minimizing the criterion becomes the function U. We introduce also the Heaviside function H and the measure of Dirac $\delta_{0}$ defined by: 
\\
\begin{center}
$H(z)=$
\begin{math}
\
\begin{array}{clrr}      
1 & if & z\leq 0\\
0  & if & z>0 \\
\end{array}\end{math}
\begin{math}
\
\begin{array}{clrr} 
et & \partial_{0}(z)=\frac{d}{dz}H(z)
\end{array}\end{math}
\\
\end{center}
The criterion is then expressed through the functions U, H and $\delta$ in the following manner:
\begin{equation}
\begin{array}{l}
J(U,c_{1},c_{2})=\int_{\Omega_{I}}\lambda\left|SV_{g}(x)-c_{1}\right|^{2}H(U(x))dx+ \\
\int_{\Omega_{I}}\lambda\left|SV_{g}(x)-c_{2}\right|^{2}(1-H(U(x)))dx+ \\
\int_{\Omega_{I}}\mu\delta(U(x))\left| \nabla U(x)\right|dx
\end{array}
\end{equation}
with:\\
\begin{equation}
\begin{array}{l}
c_{1}=\frac{\int_{\Omega}SV_{g}(x)H(U(x))dx}{\int_{\Omega}H(U(x))dx}\\
c_{2}=\frac{\int_{\Omega}SV_{g}(x)(1-H(U(x)))dx}{\int_{\Omega}(1-H(U(x)))dx}\\
\end{array}
\end{equation}
To calculate the Euler-Lagrange equation for unknown function U, we consider a regularized versions for the functions H and $\delta$ noted $H_\epsilon$ and $\delta_\epsilon$. The evolution equation is found then expressed directly with U, the function of the level set:\\

\begin{equation}
\begin{array}{l}
\frac{\partial U}{\partial \tau}=\delta_{\epsilon}(U)[\mu div (\frac{\nabla U}{\left| \nabla U \right|})+\lambda \left|SV_{g}(x)-c_{1}\right|^{2}\\- \lambda \left| SV_{g}(x)-c_{2}\right|^{2}] ( in  \Omega_{I})\\
\\
\frac{\delta_{\epsilon}(U)}{\left|\nabla U \right|}\frac{\partial U}{\partial N}=0  (on  \partial\Omega_{I})
\end{array}
\end{equation}

with $div (\frac{\nabla U}{\left| \nabla U \right|})$ the curvature of the level curve of U via x and $\frac{\partial U}{\partial N}$ the derivative of U compared to normal inside the curve N.
\subsection{Supported video surveillance concepts and events}
Until now, our system supports 5 concepts and 6 events.
The 5 concepts supported by our system are as follows:
\begin{itemize}
	\item C1: Approaching vehicule to the camera (figure 3.a)
	\item C2: One or more moving vehicule (figure 3.b)
	\item C3: Approaching pedestrian (figure 3.c)
	\item C4: One or more moving pedestrian (figure 3.d)
		\item C5: Combinated Concept (figure 3.e)
		\end{itemize}
		In our system we target six class of event. We divide this list into two categories:\\
\textbf{Collaborative events:}
\begin{itemize}
	\item Embrace
\item People Split Up
\end{itemize}
\textbf{Individual events:}
\begin{itemize}
	\item Elevator No Entry
\item	Object Put
\item	Person Runs
\item	Opposing Flow
\end{itemize}

	\begin{figure}[ht]
\begin{minipage}[b]{.30\linewidth}
  \centering
 \centerline{\epsfig{figure=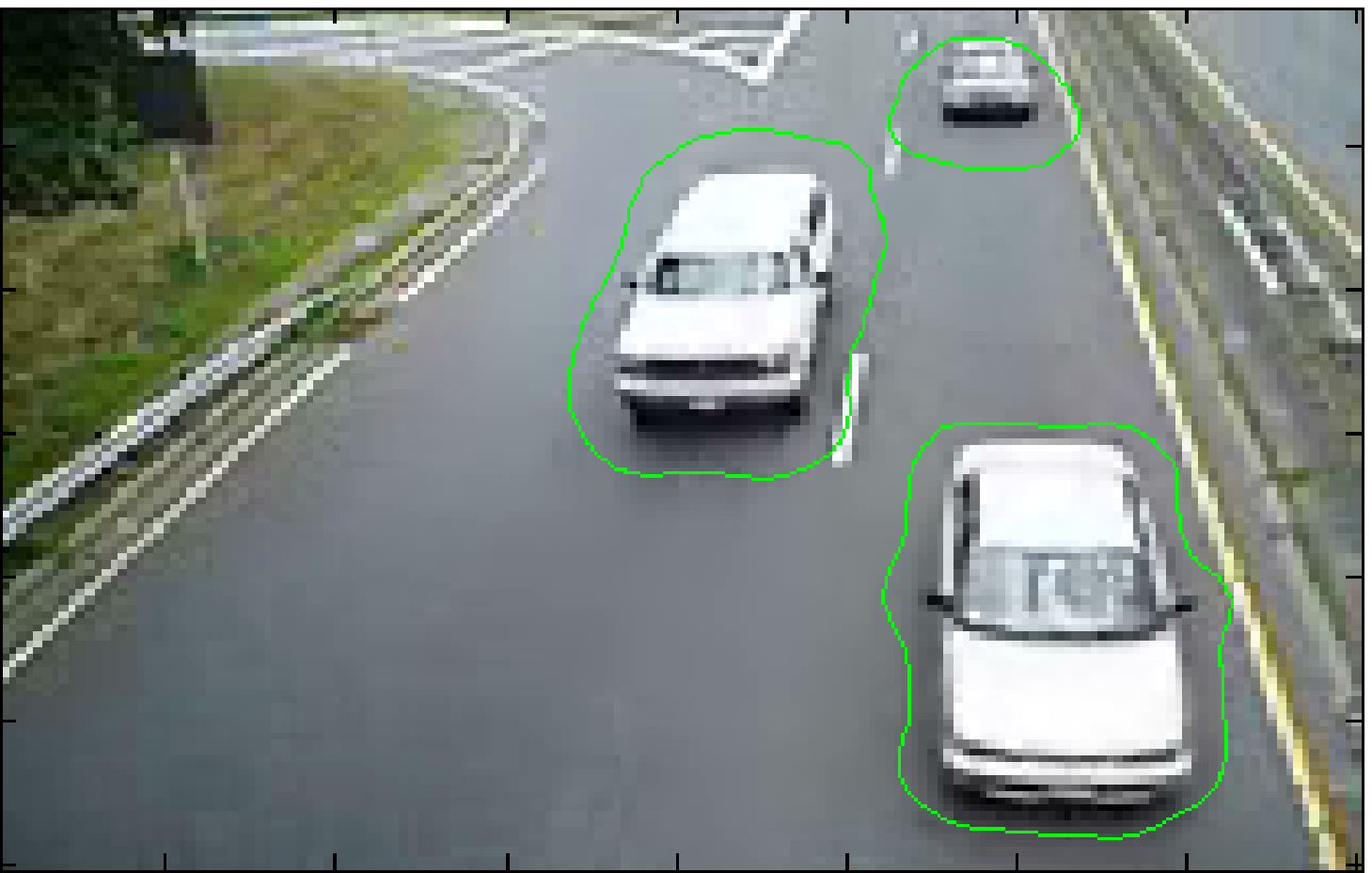,width=2.5cm}}
  \vspace{.2cm}
  \centerline{(a)}\medskip
\end{minipage}
\hfill
\begin{minipage}[b]{0.30\linewidth}
  \centering
 \centerline{\epsfig{figure=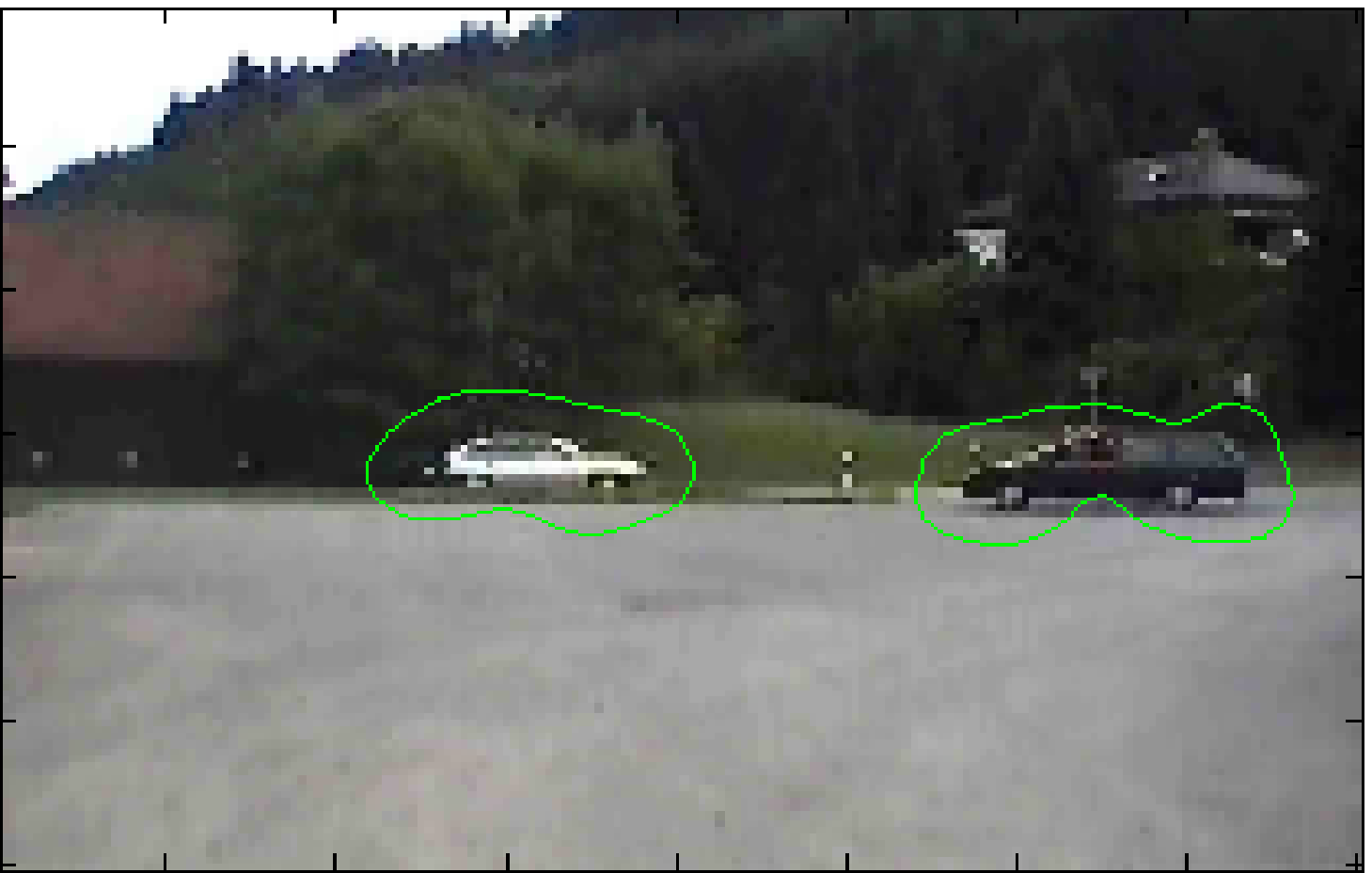,width=2.5cm}}
  \vspace{.2cm}
  \centerline{(b)}\medskip
\end{minipage}
\begin{minipage}[b]{.30\linewidth}
  \centering
 \centerline{\epsfig{figure=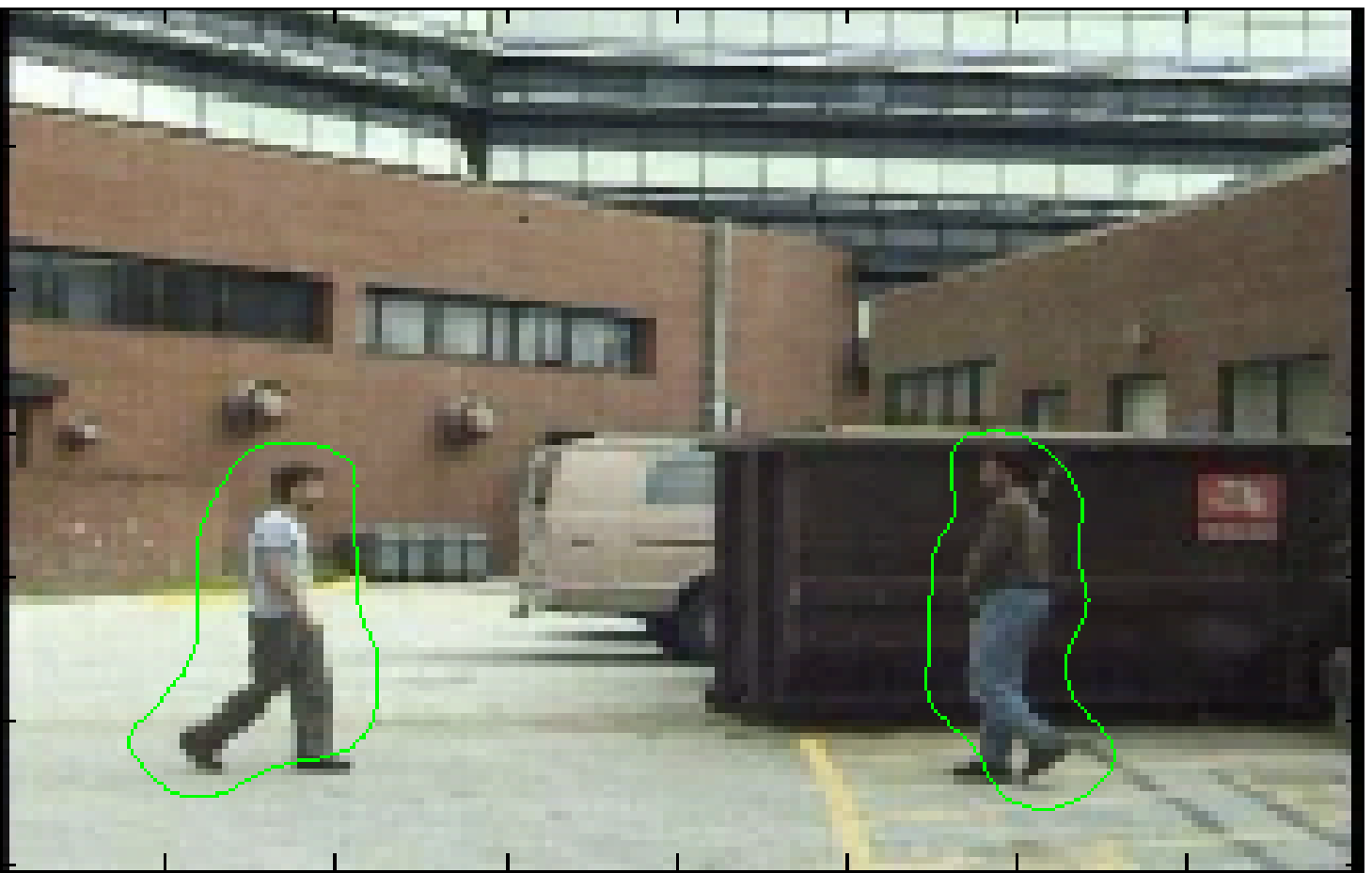,width=2.5cm}}
  \vspace{.2cm}
  \centerline{(c)}\medskip
\end{minipage}
\begin{minipage}[b]{.30\linewidth}
  \centering
 \centerline{\epsfig{figure=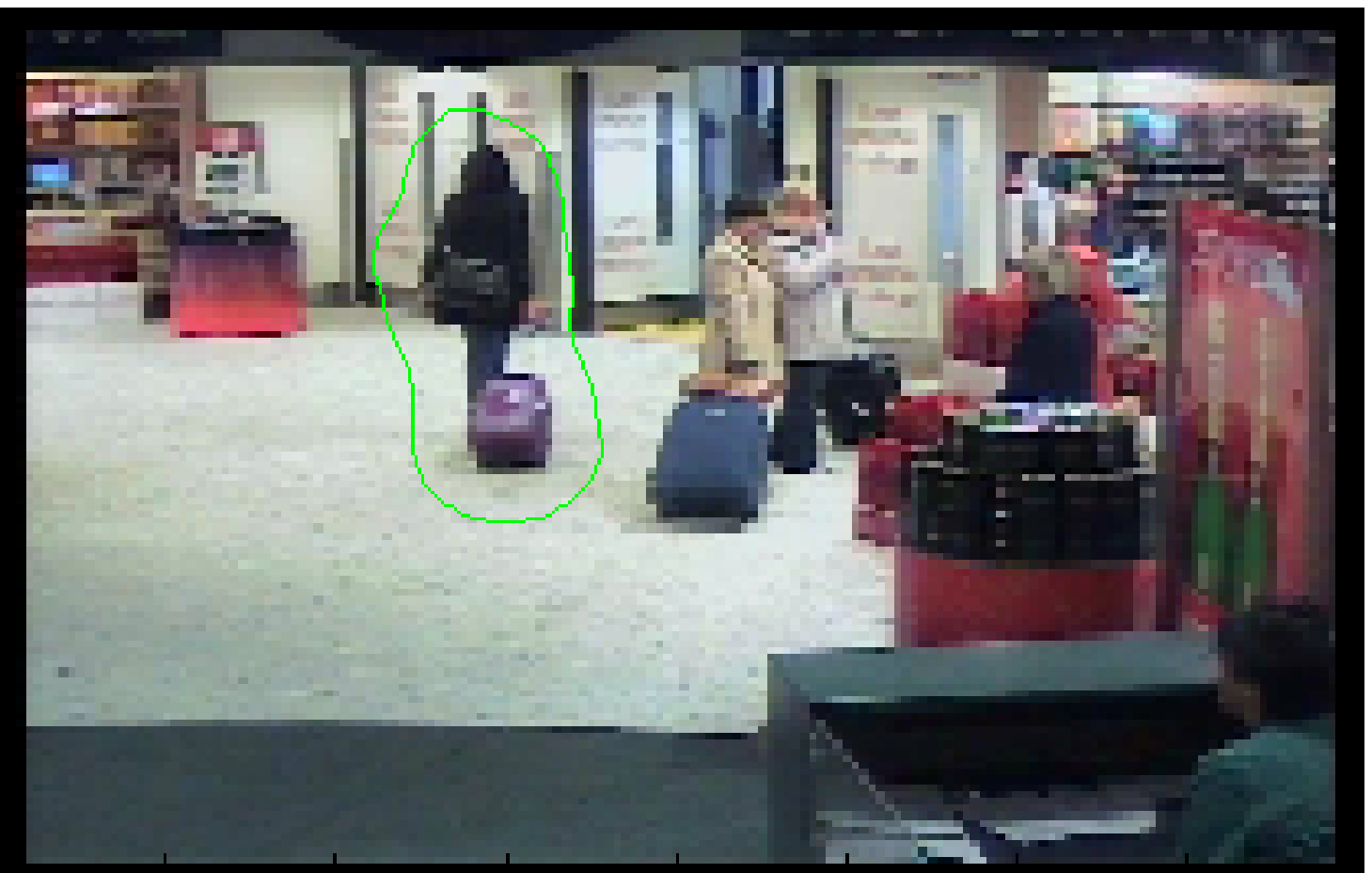,width=2.5cm}}
  \vspace{.2cm}
  \centerline{(d)}\medskip
\end{minipage}
\begin{minipage}[b]{.3\linewidth}
  \centering
 \centerline{\epsfig{figure=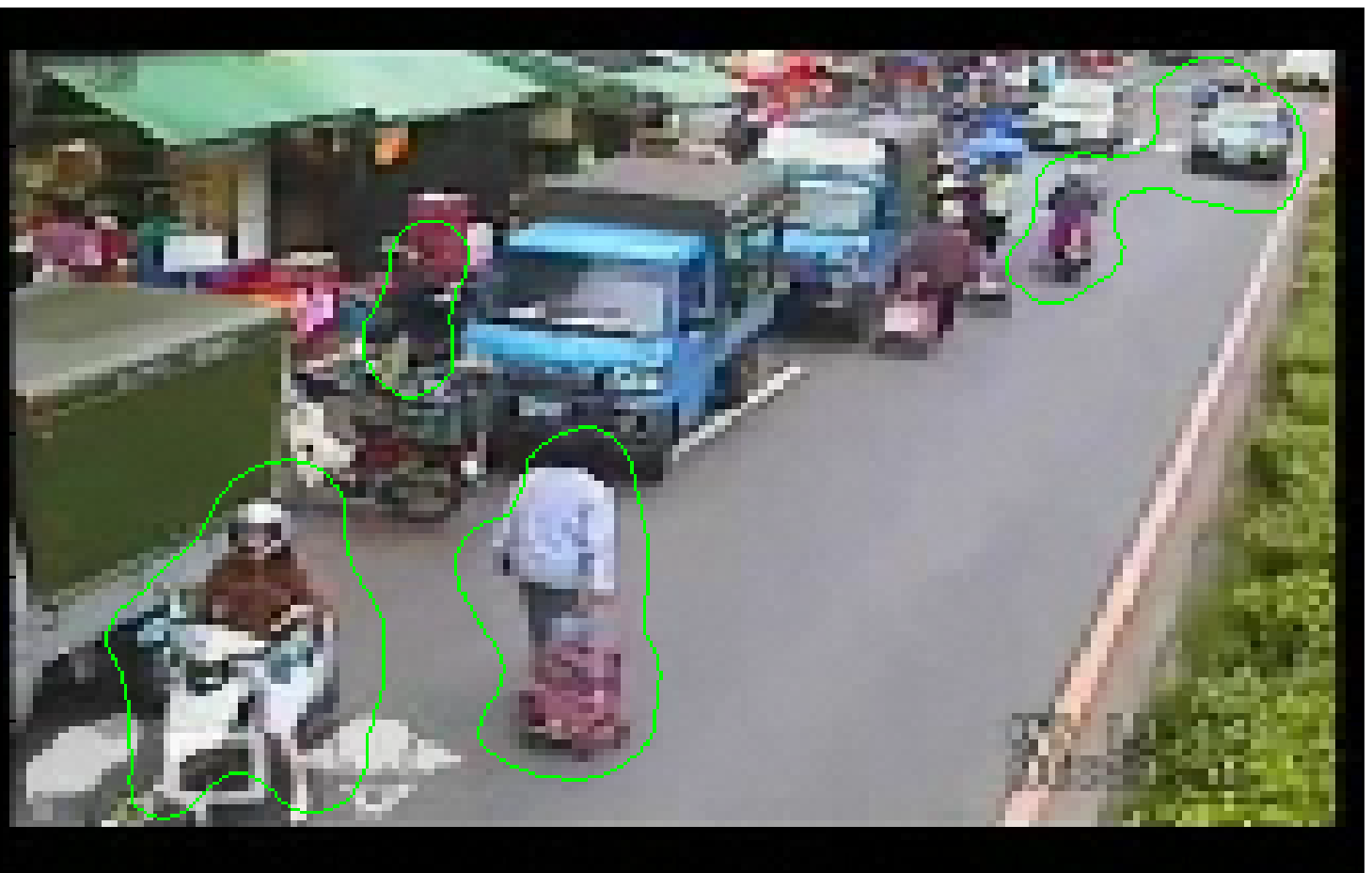,width=2.5cm}}
  \vspace{.2cm}
  \centerline{(e)}\medskip
\end{minipage}
\caption{Examples of images extracted from our video surveillance dataset.}
\label{fig:res}
\end{figure}

\begin{figure}[htb]
\begin{minipage}[b]{.3\linewidth}
  \centering
 \centerline{\epsfig{figure=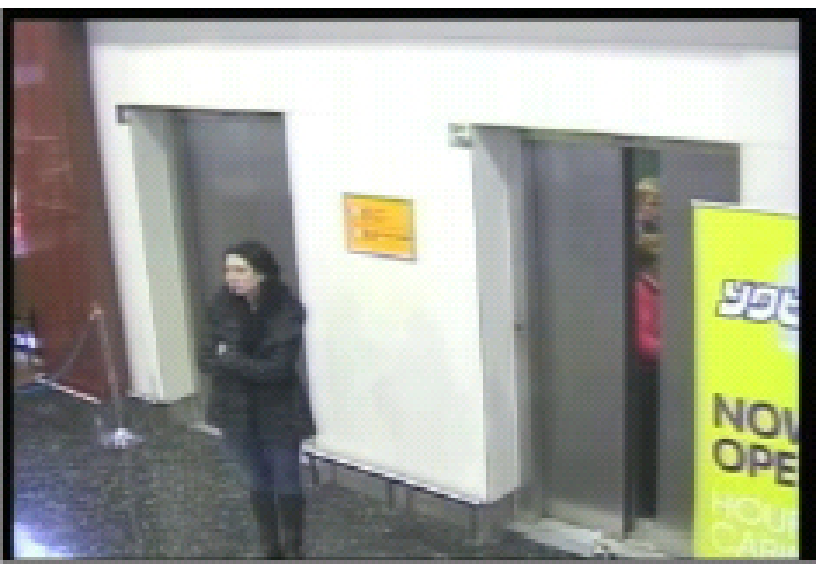,width=2.5cm}}
  \vspace{.2cm}
  \end{minipage}
\begin{minipage}[b]{.30\linewidth}
  \centering
 \centerline{\epsfig{figure=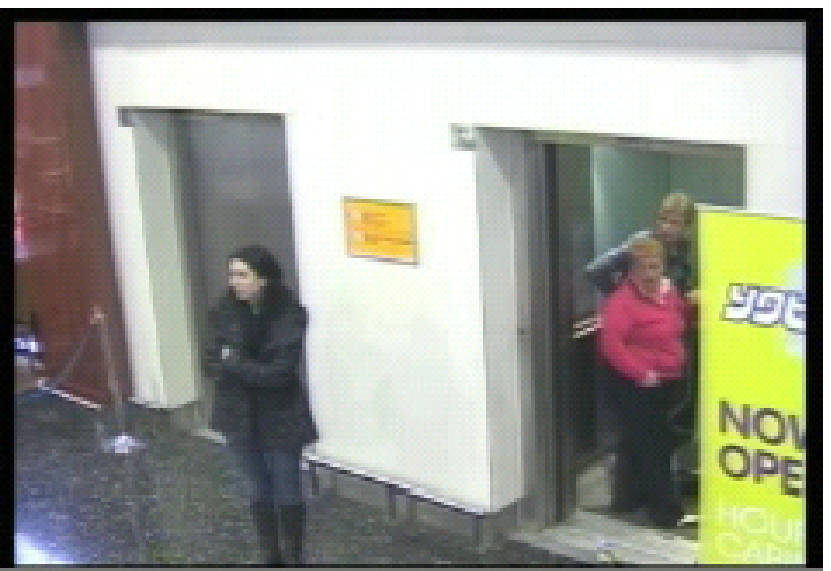,width=2.5cm}}
  \vspace{.2cm}
  \end{minipage}
\begin{minipage}[b]{.30\linewidth}
  \centering
 \centerline{\epsfig{figure=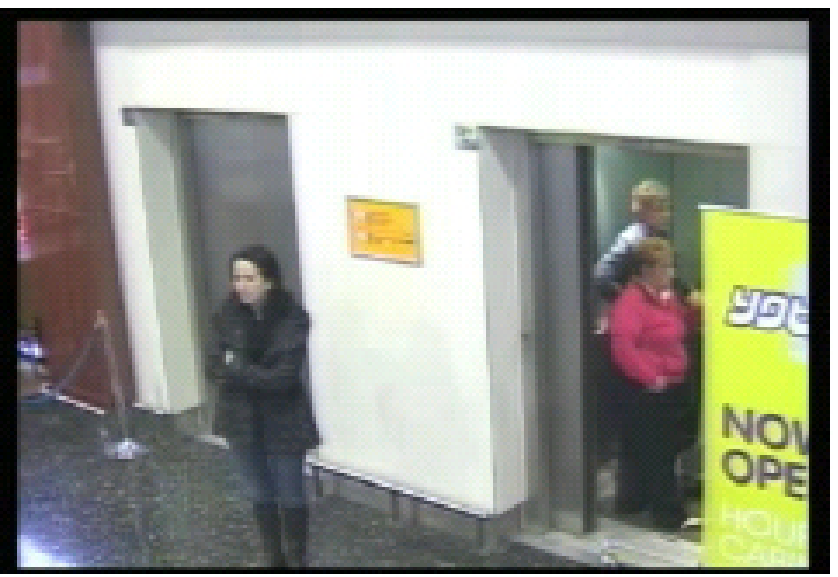,width=2.5cm}}
  \vspace{.2cm}
  \end{minipage}
  \begin{minipage}[b]{.3\linewidth}
  \centering
 \centerline{\epsfig{figure=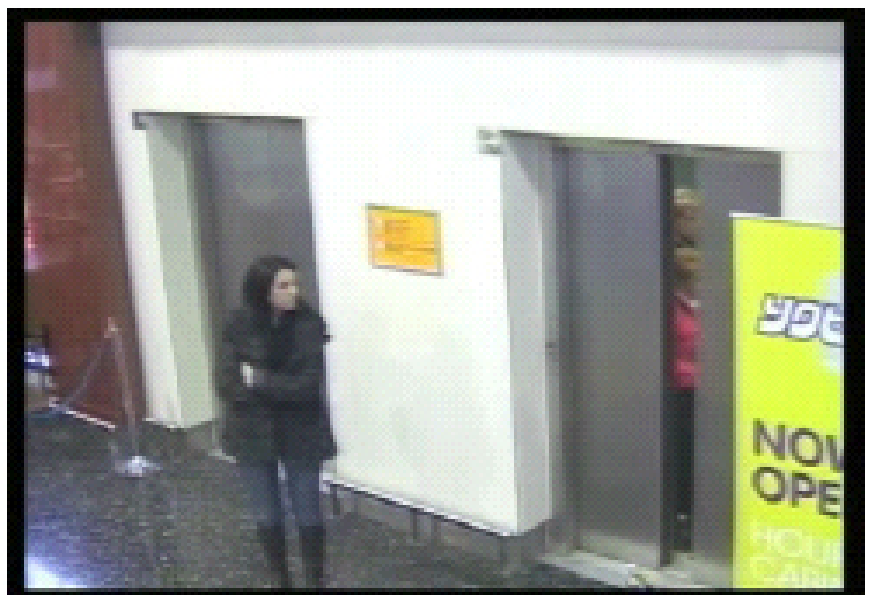,width=2.5cm}}
 \vspace{.2cm}
  
\end{minipage}
\centerline{}\medskip
 \hfill
\caption{Elevator No entry event.}
\label{fig:resfig}
\end{figure}
\begin{figure}[htb]
\begin{minipage}[b]{.3\linewidth}
  \centering
 \centerline{\epsfig{figure=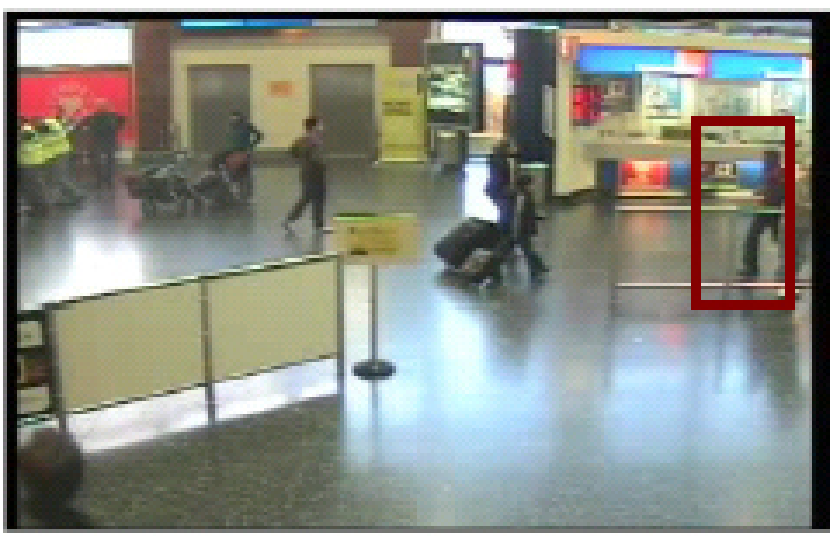,width=2.5cm}}
 \vspace{.2cm}
  
\end{minipage}
\begin{minipage}[b]{.3\linewidth}
  \centering
 \centerline{\epsfig{figure=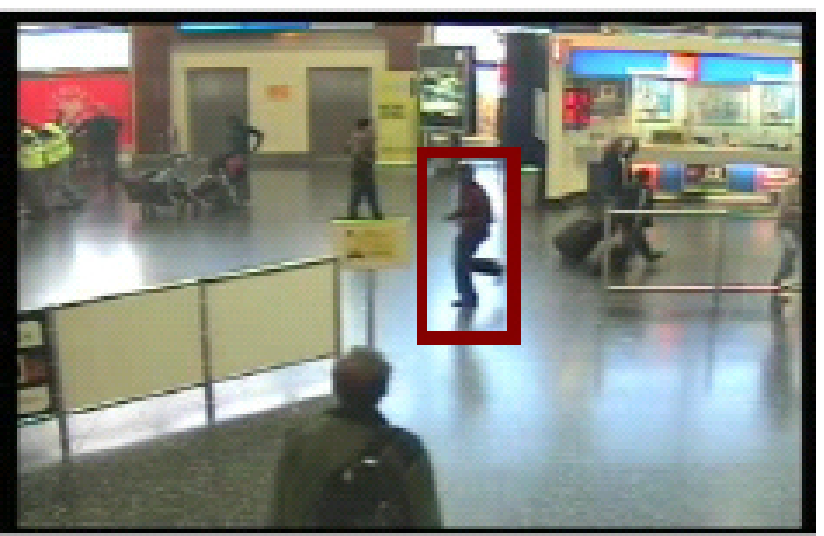,width=2.5cm}}
 \vspace{.2cm}
  \end{minipage}
  \centerline{Person Run}\medskip
\hfill
\begin{minipage}[b]{.3\linewidth}
  \centering
 \centerline{\epsfig{figure=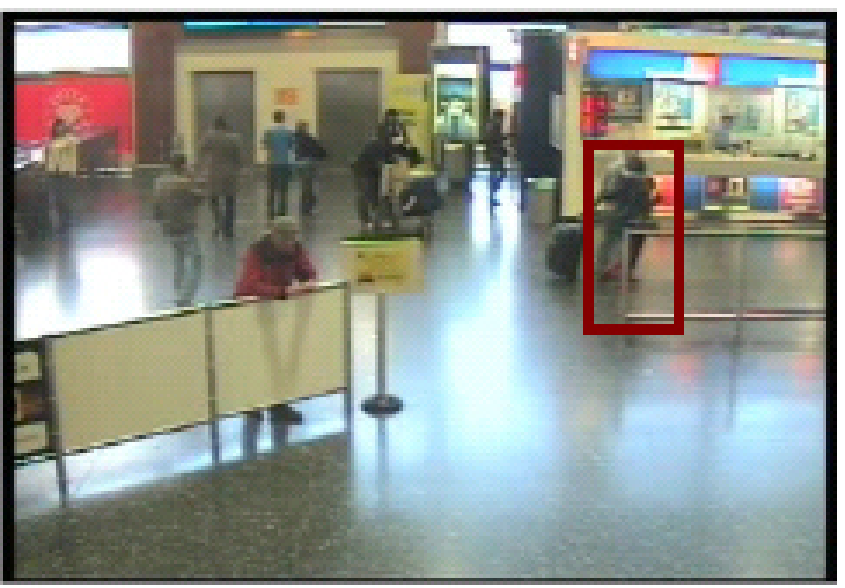,width=2.5cm}}
 \vspace{.2cm}
  
\end{minipage}
\begin{minipage}[b]{.3\linewidth}
  \centering
 \centerline{\epsfig{figure=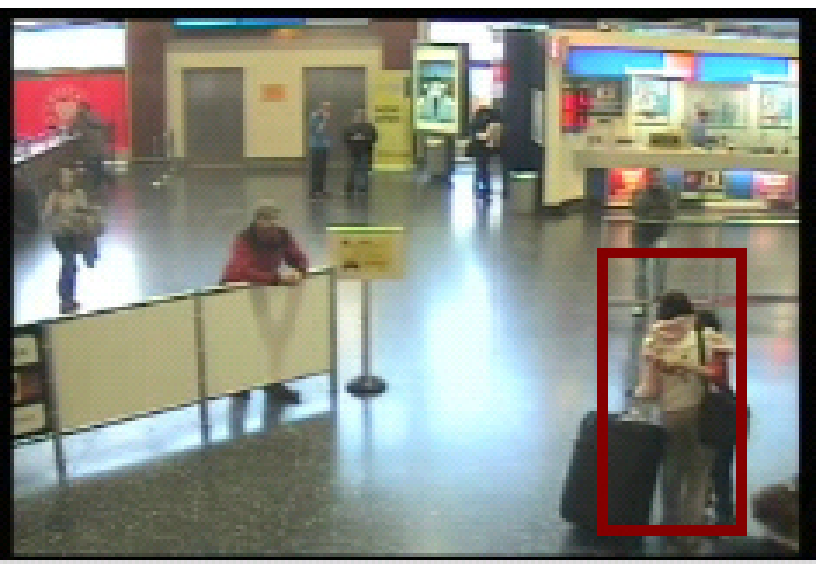,width=2.5cm}}
 \vspace{.2cm}
  \end{minipage}
  \centerline{Embrace}\medskip
\hfill
\caption{Person Run and Embrace events.}
\label{fig:resfig}
\end{figure}
\begin{figure}[htb]
\begin{minipage}[b]{.3\linewidth}
  \centering
 \centerline{\epsfig{figure=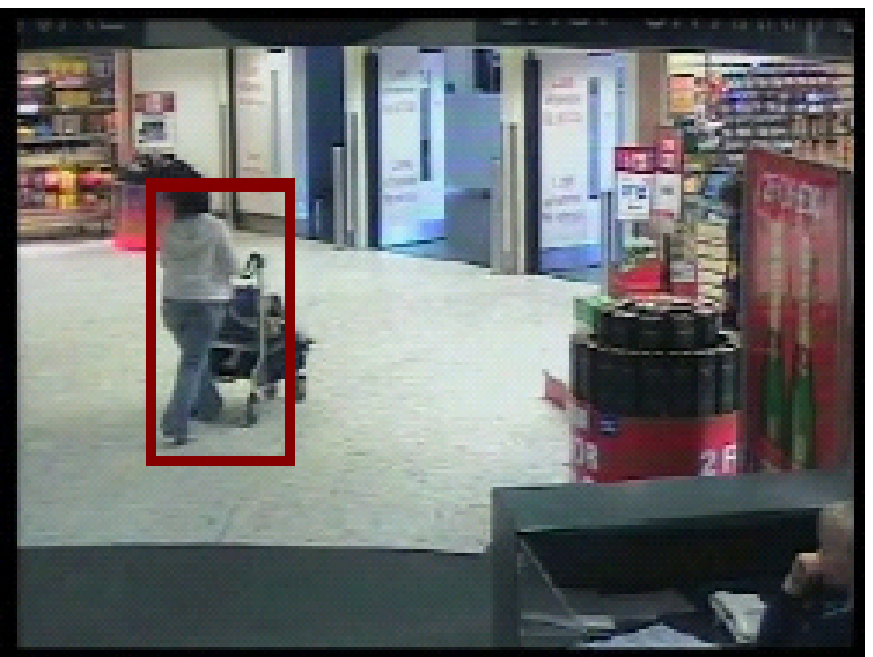,width=2.5cm}}
  \vspace{.2cm}
  \end{minipage}
\begin{minipage}[b]{.3\linewidth}
  \centering
 \centerline{\epsfig{figure=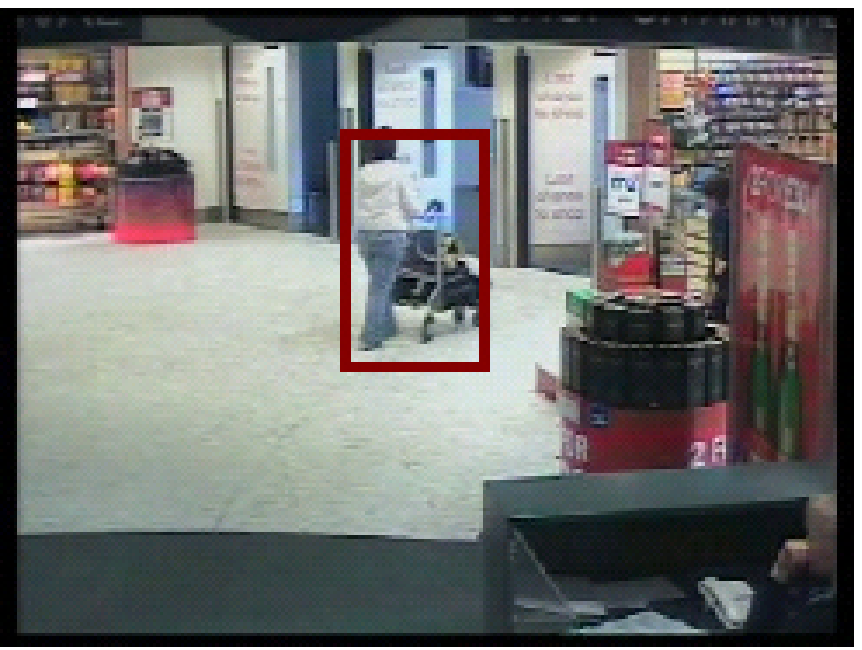,width=2.5cm}}
  \vspace{.2cm}
  \end{minipage}
\begin{minipage}[b]{.3\linewidth}
  \centering
 \centerline{\epsfig{figure=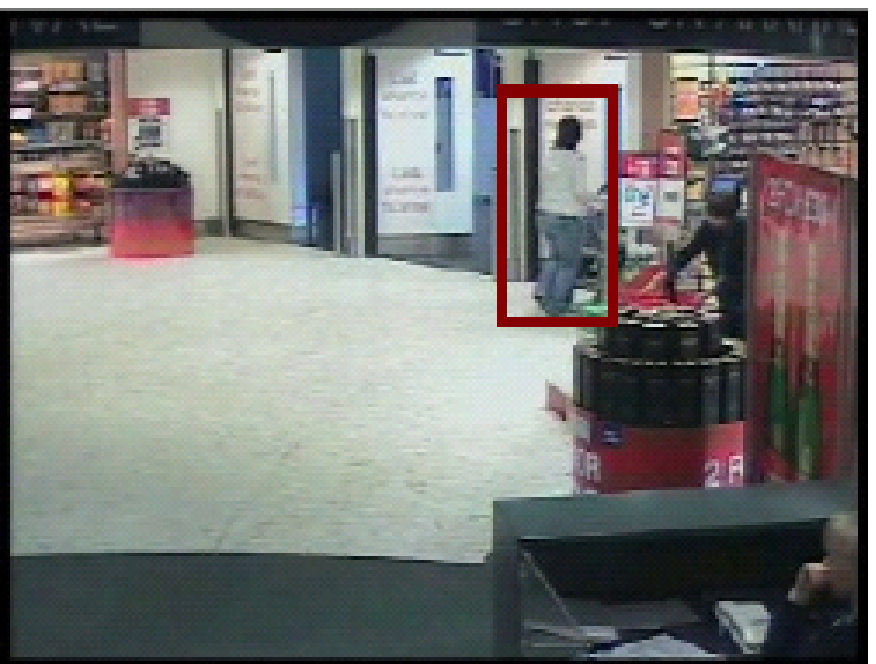,width=2.5cm}}
  \vspace{.2cm}
  \end{minipage}
  \begin{minipage}[b]{.3\linewidth}
  \centering
 \centerline{\epsfig{figure=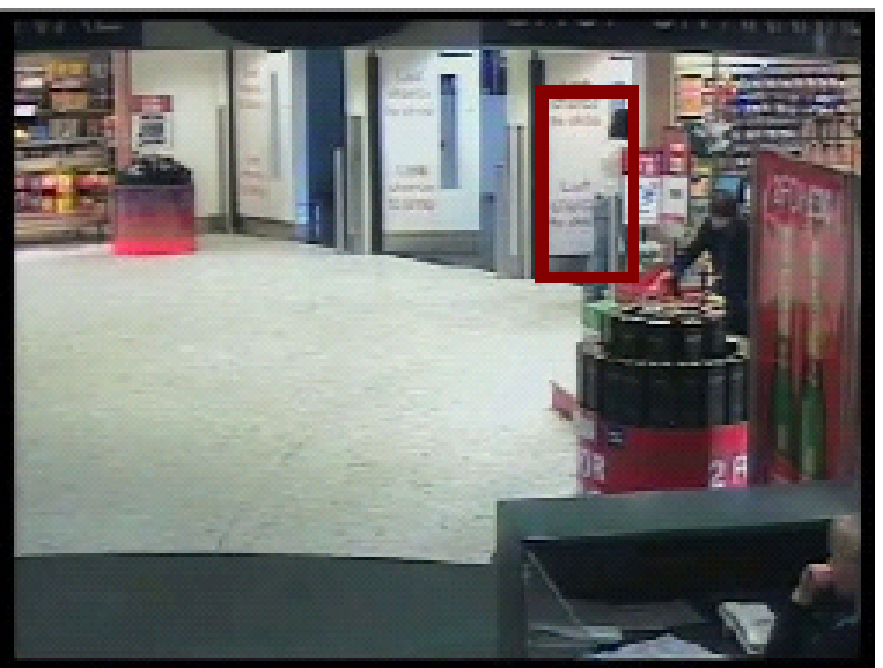,width=2.5cm}}
 \vspace{.2cm}
  
\end{minipage}
 \centerline{Opposing Flow}\medskip
\hfill
\begin{minipage}[b]{.3\linewidth}
  \centering
 \centerline{\epsfig{figure=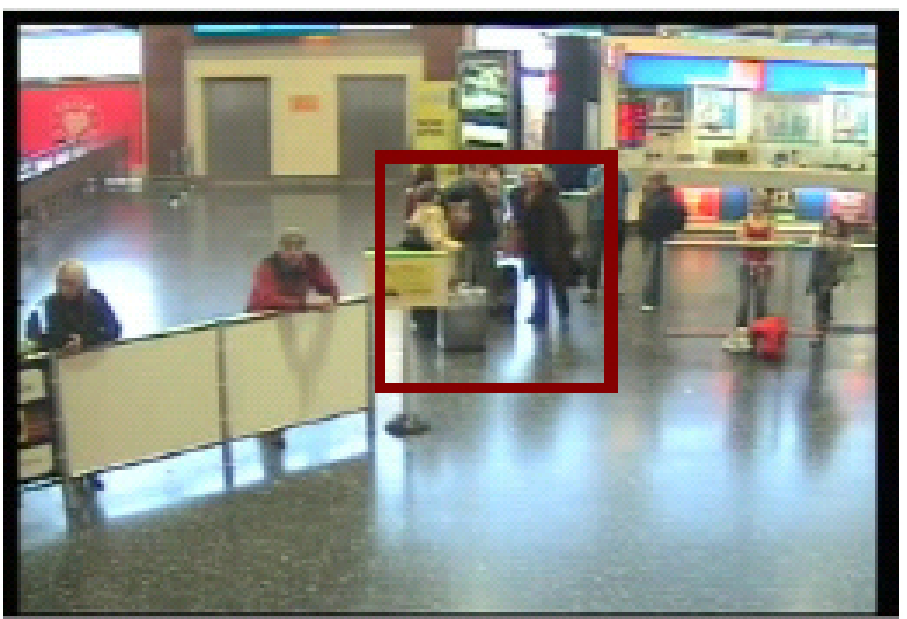,width=2.5cm}}
 \vspace{.2cm}
  
\end{minipage}
\begin{minipage}[b]{.3\linewidth}
  \centering
 \centerline{\epsfig{figure=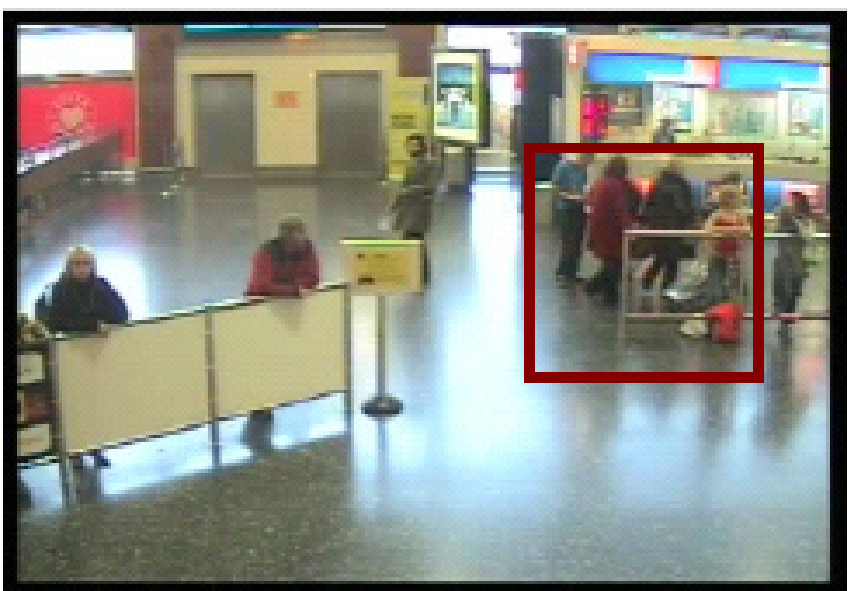,width=2.5cm}}
 \vspace{.2cm}
  \end{minipage}
  \centerline{People Split Up}\medskip
\hfill
\caption{Opposing Flow and People Split Up events.}
\label{fig:resfig}
\end{figure}

\section{Combining single SVM classifier for learning video event}

Support Vector Machines (SVMs) have been applied successfully to solve many problems of classification and regression.
However, SVMs suffer from a phenomenon called 'catastrophic forgetting', which involves loss of information learned in the presence of new training data. Learn++ \cite{bib7} has recently been introduced as an incremental learning algorithm. The strength of Learn++ is its ability to learn new data without forgetting prior knowledge and without requiring access to any data already seen, even if new data introduce new classes. To benefit from the speed of SVMs and the ability of incremental learning of Learn++, we propose to use a set of trained classifiers with SVMs based on Learn++ inspired from \cite{bib8}.
Experimental results of detection of events suggest that the proposed combination is promising. 
According to the data, the performance of SVMs is similar or even superior to that of a neural network or a Gaussian mixture model.
\subsection{SVM Classifier}
Support Vector Machines (SVMs) are a set of supervised learning techniques to solve problems of discrimination and regression. The SVM is a generalization of linear classifiers.%
The SVMs have been applied to many fields (bio-informatics, information retrieval, computer vision, finance ...). \\
According to the data, the performance of SVMs is similar or even superior to that of a neural network or a Gaussian mixture model.
They directly implement the principle of structural risk minimization \cite{bib9} and work by mapping the training points into a high dimensional feature space, where a separating hyperplane $(w, b)$ is found by maximizing the distance from the closest data points (boundary-optimization). Given a set of training samples $ S=\left\{(x_i,y_i) | i=1,..,m\right\}$, where $x_i \in R_n$ are input patterns, $y_i  \in {+1, -1}$ are class labels for a 2-class problem, SVMs attempt to find a classifier $h(x)$, which minimizes the expected misclassification rate. A linear classifier $h(x)$ is a hyperplane, and can be represented as $h(x)=sign(w^{T}x+b)$. The optimal SVM classifier can then be found by solving a convex quadratic optimization problem:
\begin{equation}
\begin{array}{l}
\underbrace{max}_{w,b}\frac{1}{2}\left\|w\right\|^2+C\sum^{m}_{i=1}\xi_{i}\ subject\ to\ \\
 y_i\left(\left\langle w,x_i\right\rangle+b\right)\geq1-\xi_i\ and\ \xi_i\geq0
\end{array}
\end{equation}
Where $b$ is the bias, $w$ is weight vector, and $C$ is the regularization parameter, used to balance the classifier's complexity and classification accuracy on the training set $S$.
Simply replacing the involved vector inner-product with a non-linear kernel function converts linear SVM into a more flexible non-linear classifier, which is the essence of the famous kernel trick. In this case, the quadratic problem is generally solved through its dual formulation:
\begin{equation}
\begin{array}{l}
L\left(w,b,\alpha\right)=\sum^{m}_{i=1}\alpha_i-\frac{1}{2}\left(\sum^{m}_{i=1}y_{i}y_{j}\alpha_i\alpha_jK\left(x_i,x_j\right)\right)\\\\
subject\ to\ C\geq\alpha_i\geq0\ and\ \sum^{m}_{i=1}y_{i}\alpha_iy_i=0
\end{array}
\end{equation}
where $a_i$ are the coefficients that are maximized by Lagrangian. For training samples $x_i$, for which the functional margin is one (and hence lie closest to the hyperplane), $\alpha_i\succ0$. Only these instances are involved in the weight vector, and hence are called the support vectors \cite{bib10}. The non-linear SVM classification function (optimum separating hyperplane) is then formulated in terms of these kernels as:
\begin{equation}
h\left(x\right)=sign\left(\sum^{m}_{i=1}\alpha_iy_{i}K\left(x_i,x_j\right)-b\right)
\end{equation}
\subsection{M-SVM Classifiers}

M-SVM is based on Learn++ algorithm. This latter, generates a number of weak classifiers from a data set with known label. Depending on the errors of the classifier generated low, the algorithm modifies the distribution of elements in the subset according to strengthen the presence of the most difficult to classify. This procedure is then repeated with a different set of data from the same dataset and new classifiers are generated. By combining their outputs according to the scheme of majority voting Littlestone we obtain the final classification rule.\\
The weak classifiers are classifiers that provide a rough estimate - about 50\% or more correct classification - a rule of decision because they must be very quick to generate.
A strong classifier from the majority of his time training to refine his decision criteria. Finding a weak classifier is not a trivial problem and the complexity of the task increases with the number of different classes, however, the use of NN algorithms can correctly resolved effectively circumvent the problem. The error is calculated by the equation:

\begin{equation}
error_t=\sum_{i:h_i\left(x_i\right)\neq y_i}S_t\left(i\right)\left[\left|h_t \left(x_i\right)\neq y_i \right|\right]
\end{equation}
with $h_t : X\rightarrow Y$ an hypothesis and   where $TR_t$ is the subset of training subset and the $TE_t$ is the test subset.
The synaptic coefficients are updated using the following equation:
\begin{equation}
w_{t+1}\left(i\right)=w_t\left(i\right)*
\left\{
\begin{array}{l}
	\beta_{t}\ if\ H_t\left(x_i\right)=y_i\\
	1\ else
	\end{array}
	\right\}
\end{equation}
Where $t$ is the iteration number, $B_t$ composite error and standard composite hypothesis $H_t$.

\begin{figure}[ht]

\epsfig{figure=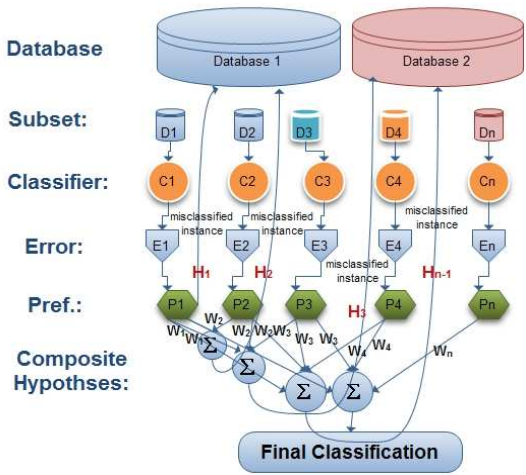,width=7cm}

   \caption{M-SVM classifier}

\end{figure}

In our approach we replace each weak classifier by SVM.  After $T_k$ classifiers are generated for each $D_k$, the final ensemble of SVMs is obtained by the weighted majority of all composite SVMs:

\begin{equation}
H_{final}\left(x\right)=arg\ \underbrace{max}_{y\in Y}\sum^{K}_{k=1}\sum_{t:h_t\left(x\right)=y}log\frac{1}{\beta_t}
\end{equation}

\section{Visual Feature Extraction}
We use a set of different visual descriptors at various granularities for each frame, rid of the static background, of the video shots. The relative performance of the specific features within a given feature modality is shown to be consistent across all concepts/events. However, the relative importance of one feature modality vs. another may change from one concept/event to the other.
The following descriptors had the top overall performance for both search and concept modeling experiments:
\begin{itemize}
	\item Color Histogram: global color represented as 128-dimensional histogram in HSV color space.
	\item Color Moments: localized color extracted from 3x3 grid and represented by the first 3 moments for each grid region in Lab color space as normalized 255-dimensional vector.
	\item Co-occurence Texture: global texture represented as a normalized 96-dimentional vector of entropy, energy, contrast and homogeneity extracted from the image gray-scale co-occurence matrix at 24 orientation.
	\item Gabor Texture: Gabor functions are Gaussians modulated by complex sinisoids. The Gabor filter masks can be considred as orientation and scale-tunable and line detectors. The statistics of these micro-features in a given region can be used to characterize the underlying texture information. We take 4 scales and 6 orientations of Gabor textures and further use their mean and standard deviation to represent the whole frame and result in 48 textures.
	\item Fourier: Features based on the Fourier transform of the binarized edge image. The 2-
dimensional amplitude spectrum is smoothed and down-sampled to form a feature
vector of 512 parameters.
	\item Sift:The SIFT descriptor \cite{David06} is consistently among the best performing interest region descriptors. SIFT describes the local shape of the interest region using edge histograms.
To make the descriptor invariant, while retaining some positional information, the interest region is divided into a 4x4 grid and every sector has its own edge direction histogram (8 bins). The grid is aligned with the dominant direction of the edges in the interest region to make the descriptor rotation invariant.
	\item Combined Sift and Gabor.
	\item Wavelet Transform for texture descriptor: Wavelets are hybrids that are waves within a region of the image, but otherwise particles. Another important distinction is between particles that have place tokens and those that do not. Although all particles have places in the image, it does not follow these places will be represented by tokens in feature space. It is entirely feasible to describe some images as a set of particles, of unknown position. Something like this
happens in many description of texture. We performe 3 levels of a Daubechies wavelet \cite{I.Daubechies1991} decomposition for each frame and calculate the energy level for each scale, which resulted in 10 bins features data.
	\item Hough Transform: As descriptor of shape we employ a histogram based on the calculation of Hough transform \cite{Nozha02}.  This histogram gives information better than those given by the edge histogram. We obtain a combination of behavior of the pixels in the image along the straight lines. 
	\item Motion Activity:We use the information calculated by the optical flow, through concentrating on movements of the various objects (people or vehicle) detected by the method described in the previous section. The descriptors that we use are correspond to the energie calculated on every sub-band, by a decomposition in wavelet of the optical flow estimated between every image of the sequence. We obtain a vector of 10 bins, they represent for every image a measure of activity sensitive to the amplitude, the scale and the orientation of the movements in the shot.
	\end{itemize}
	\section{Experimental Results}
	Experiments are conducted on the many sequence from TRECVid'2009 database of video surveillance and many other sequences from road traffics. About 20 hours are used to train the feature extraction system, that are segmented in the shots. These shots were annotated with items in a list of 5 concepts and 6 events. We use about 20 hours for the evaluation purpose.\\
	To evaluate the performance of our concept detection sub-system, we use the common measure from the information retrival community: the Average Precision.
	Figure 8 shows the evaluation of returned shots. The best results are obtained for concepts: 1,2,3,5. The remaining run also provide satisfying results.

	\begin{table}[ht]\footnotesize
	\centering
	\caption{Event detection Results}
\begin{tabular}{|c|c|c|c|c|}
\hline
\hline 
 & \multicolumn{2}{c|}{Run 1} & \multicolumn{2}{c|}{Run 2}\tabularnewline

\hline 
 \hline 
\textbf{Event} & \textbf{A.NDCR} & \textbf{M.NDCR} & \textbf{A.NDCR} & \textbf{M.NDCR}\tabularnewline
\hline 
Embrace & 1.045 &  0.981 & 1.025 & 0.961\tabularnewline
\hline 
PeopleSplitUp & 3.432 & 0.983 & 3.331 & 0.953\tabularnewline
\hline 
ElevatorNoEntry & 0.341 & 0.332 & 0.331 & 0.322\tabularnewline
\hline 
ObjectPut  & 1.076 & 0.997 & 1.056 & 0.952\tabularnewline
\hline 
PersonRuns & 1.087 & 0.976 & 1.016 & 0.956\tabularnewline
\hline 
OpposingFlow & 1.032 & 1.021  & 0.993 & 0.923\tabularnewline
\hline
\end{tabular}
\end{table}
To evaluate the performance of event detection sub-system we use the TRECVID'2009 event detection metrics. The evaluation uses the Normalized Detection Cost Rate (NDCR).  \\ 
NDCR is a weighted linear combination of the system's Missed Detection Probability and False Alarm Rate (measured per unit time).  The measure's derivation can be found in ($\small {http://www.itl.nist.gov/iad/mig/tests/trecvid/2009}$\\ 
$/doc/EventDet09-EvalPlan-v03.htm$) and the final formula is summarized below.
 Two versions of the NDCR will be calculated for the system: the Actual NDCR and the Minimum NDCR.  \\
The actual and minimum NDCRs for each of the events can be seen in Table 1. We have achieved very competitive minimum DCR results on the events of embrace, people Split UP, Object Put, opposing Flow and especially for Elevator No Entry. We did not extensively tune parameters with the aim of producing low actual DCR score; our actual DCR looks relatively higher (the lower the score, the better the performance). But our system achieved very good minimum DCR scores.

\begin{figure}[htb]
  \centerline{\epsfig{figure=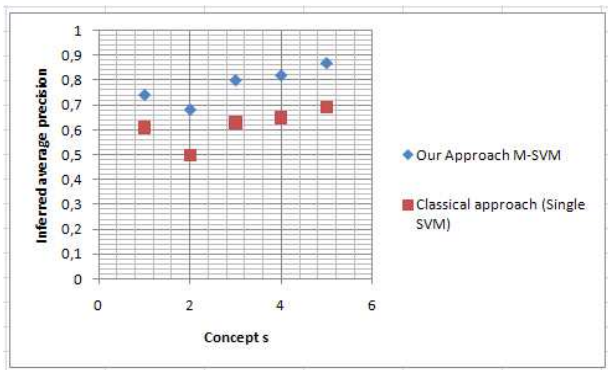,width=8cm}}
  \caption{Our run score versus Classical System (Single SVM) by Concept.}
	\label{fig:Drawing1}
  \end{figure}
\section{Conclusion}
\label{Sec:Con}
In this paper, we have presented a new approach for high-level feature extraction for video surveillance indexing and retrieval. The results obtained so far are interesting and promoters.The advantage of our approach is that allows human operators to use context-based queries and the response to these queries is much faster.
The meta-data layer allows the extraction of the motion and objects descriptors from video key-frames to XML files that then can be used by external applications such as multimedia data mining systems. Finally, the system functionalities will be enhanced by a complementary tools to improve the basic concepts and events taken care of by our system.

\section{Acknowledgement}
The authors would like to acknowledge the financial support of this work by grants from General Direction of Scientific Research (DGRST), Tunisia, under the ARUB program.
\bibliographystyle{plain}
\bibliography{files/icgst}

\end{document}